\documentclass[showpacs,amssymb,twocolumn,aps]{revtex4}
\usepackage{amssymb}
\usepackage{amsmath}
\usepackage[pdftex]{graphicx}
\usepackage{epstopdf}
\usepackage[utf8]{inputenc}
\usepackage{fancyhdr}
\usepackage{slashed}

\begin{document}

\date{\today}
\title{Nonanalyticity of the free energy in thermal field theory}
\author{F. T. Brandt\footnote[2]{fbrandt@usp.br}, J. Frenkel\footnote[3]{jfrenkel@fma.if.usp.br}
and J. B. Siqueira\footnote[8]{joao@fma.if.usp.br} }
\affiliation{ Instituto de F\'{\i}sica,
Universidade de S\~ao Paulo,
S\~ao Paulo, SP 05315-970, Brazil}
\begin{abstract}
We study, in a $d$-dimensional space-time, the nonanalyticity of the thermal free energy in the scalar $\phi^4$ theory as well as in QED. 
We find that the infrared divergent contributions induce, when $d$ is even, a nonanalyticity in the coupling $\alpha$ of the form $(\alpha)^{(d-1)/2}$ 
whereas when $d$ is odd the nonanalyticity is only logarithmic.
\end{abstract}


\maketitle

As is well known, the presence of infrared divergences in field
theories at finite temperature leads to a breakdown of the 
na\"{\i}ve perturbation theory \cite{kapusta:book89,lebellac:book96}. 
Then, a necessary first step is a resummation of the hard thermal loops, which leads to an effective perturbative expansion \cite{Frenkel:1990br,Braaten:1990mz}. 
In this context, a physical quantity of interest is the free energy of a system of hot quanta and particles, in thermal equilibrium. This problem has been much studied in four-dimensional space-time, 
where contributions to the free energy have been evaluated to higher loop order \cite{frenkel:1992az,Arnold:1994ps,Kajantie:2000iz,Blaizot:2003iq,Vuorinen:2002ue,Brandt:2006bf}. 
In this case, although the individual infrared divergent contributions add up to an IR convergent result in the resummed theory, these have the effect of introducing  a nonanalyticity in the perturbative series. 
This is manifested firstly through the appearance of terms of order $\alpha^{3/2}$, where $\alpha$ is the effective coupling constant.

Our work is motivated by the fact that field theories in $d\neq 4$ dimensional space-time may also be relevant from a physical point of view. For example, important phenomena like the high-$T_c$ superconductivity or the
fractional quantum Hall effect can be understood in the framework of QED in lower dimensions  \cite{zee:book03}. On the other hand, in higher dimensions, simple scalar theories with self-coupled spinless fields exhibit
some interesting similarities with the gauge theories of QCD  \cite{muta:book87} or quantum gravity \cite{Altherr:1991fu,Brandt:2008gy}. 

The purpose of this note is to study the lowest order nonanalyticity in the free energy of  thermal field theories defined in a $d$-dimensional space-time. 
We find that there is a significant difference in the behavior of the free energy, according to whether the dimensionality $d$ of space-time is even or odd. 
We  show that, when $d$ is even, the free energy exhibits a powerlike nonanalyticity, of order $\alpha^{(d-1)/2}$. 
On the other hand, in a space-time where $d$ is odd, the nonanalyticity is only logarithmic, behaving like $\ln \alpha$. In order to explain this behavior, we  first discuss, 
for more clarity, a scalar field theory with an interaction potential $g^2 \phi^4$. As we will see, the basic features exhibited by the free energy in this theory may then be readily extended 
to QED.

\begin{figure}
\includegraphics[scale=0.45]{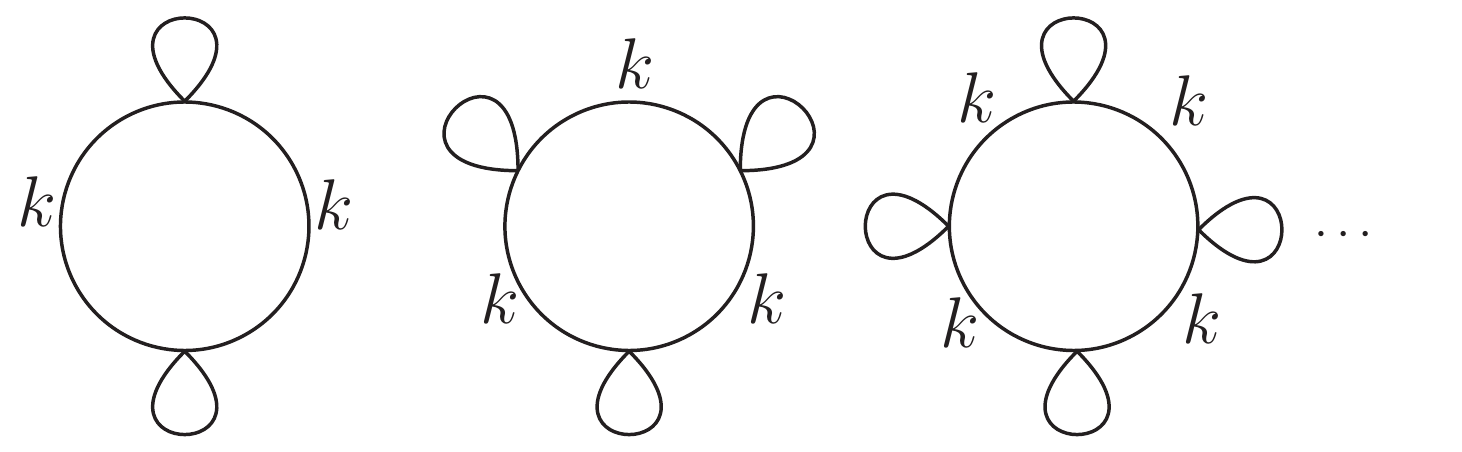}
\caption{Ring diagrams in the $\phi^4$ model. Small blobs represent one-loop self-energy diagrams.}
\label{rings-scalar}
\end{figure}

For our purpose, we find convenient to work in the imaginary time formalism \cite{kapusta:book89,lebellac:book96}, where the energies take the discrete values $\omega_n = 2 \pi n\, T$. 
We will assume that the temperature $T$ is much higher than the zero-temperature masses, which consequently will be neglected. 
It can easily be seen that the infrared divergences come only from the static mode $n=0$. For the free energy in the scalar model, the resummation method leads to the so-called rings diagrams shown 
in Fig.~\ref{rings-scalar}.

In a $d$-dimensional space-time, we find that the one-loop thermal self-energy of the scalar particle has at high temperature a leading behavior given by
\begin{align}
\Pi_\beta =  g^2 T^{d-2} \frac{ \Gamma(d-2) \zeta(d-2)}{2^{d-1} \pi^{(d-1)/2} \Gamma\left( \frac{d-1}{2} \right) },
\label{eq:pipahi}
\end{align} 
where $\Gamma$ and $\zeta$ denote, respectively, the gamma and the Riemann zeta functions \cite{gradshteyn}.
By simple power counting, it is easy to verify that the infrared contributions start when the number of self-energy loops is $N=d/2$, for $d$ even, or $N=(d-1)/2$ for $d$ odd. 
We denote by $E(d/2)$ the integer part of $d/2$, so that  $E(d/2)=d/2$ or $E(d/2)=(d-1)/2$ respectively for even or odd $d$. 
We then obtain, from the dominant infrared divergent terms of the ring diagrams, the following contributions to the free-energy
\begin{align}
\Omega_{ring}(T) = \frac{1}{2} VT \sum_{N= E(d/2)}^{\infty} \int \frac{d^{d-1} k}{( 2 \pi)^{d-1}} \frac{(-1)^{N+1}}{N} \left( \frac{\Pi_\beta}{\underline{k}^2} \right)^N ,
\label{eq:omegaphi41}
\end{align}  
where $V$ is the ($d-1$)-dimensional volume, $1/N$ is the symmetry factor and $\underline{k}$ is the ring momentum.  
One can see that when $d$ is even the sum begins with a linearly infrared divergent contribution, 
which becomes stronger as $N$ increases. On the other hand, when $d$ is odd the sum begins with a logarithmically infrared divergent contribution, so that in this case the infrared singularities 
are weaker termwise.

Performing the summation in Eq.~\eqref{eq:omegaphi41}, we obtain
\begin{align}
\Omega_{ring}(T) &= \frac{1}{2} VT  \int \frac{d^{d-1} k}{( 2 \pi)^{d-1}} \left[ \ln\left( 1 + \frac{\Pi_\beta}{\underline{k}^2} \right)  \right. \nonumber \\ &- \left. \sum_{N=1}^{E(d/2)-1} \frac{(-1)^{N+1}}{N} \left( \frac{\Pi_\beta}{\underline{k}^2}  \right)^N \right],
\label{eq:omegaphi42}
\end{align}
where the last term should be omitted for $d=2$ or $d=3$.
Although this result is IR convergent, it does exhibit a nonanalyticity in the coupling constant $g^2$ present in $\Pi_\beta$. 
To see this, it will be convenient to study separately the cases when $d$ is even or odd, as it turns out that the nonanalyticity behaves differently in these cases.

When $d$ is even, we may perform an integration by parts in \eqref{eq:omegaphi42} so that, since the surface terms vanish, one finds
\begin{align}
\Omega^{d-even}_{ring}(T) &=  VT  \frac{(-1)^{d/2+1}}{2^d \pi^{(d-3)/2} \Gamma \left(\frac{d+1}{2} \right)}    \left( \Pi_\beta \right)^{(d-1)/2} 
\label{eq:omegaphi4even} 
\end{align}
where the factor $\left( \Pi_\beta \right)^{(d-1)/2}$ basically arises on dimensional grounds. 
We see that, since $d$ is even, $\frac{d-1}{2}$ is a half-integer, so \eqref{eq:omegaphi4even} exhibits a powerlike nonanalyticity in the coupling constant $g^2$. 
When $d=4$, \eqref{eq:omegaphi4even} reduces to the well-known result \cite{kapusta:book89,lebellac:book96}
\begin{align}
\Omega^{d=4}_{ring}(T) = - \frac{VT^4}{12\pi} \left(\frac{g^2}{24} \right)^{3/2}.
\end{align}

On the other hand, when $d$ is odd, one can no longer neglect the surface term when \eqref{eq:omegaphi42} is integrated by parts.
This term, which arises from the region $|\underline{k}|\rightarrow \infty$, turns out to give the nonvanishing contribution
\begin{align}\label{surface}
&\frac{VT}{(2\pi)^{d-1}}\frac{(-1)^{\frac{d+1}{2}}}{(d-1)^2} \frac{2\pi^{\frac{d-1}{2}}}{\Gamma\left(\frac{d-1}{2}\right)}|\underline{k}|^{d-1} \left(\frac{\Pi_\beta}{\underline{k}^2}\right)^{\frac{d-1}{2}}
\nonumber \\ & =
\frac{VT}{d-1}\frac{(-1)^{\frac{d+1}{2}}}{2^{(d-1)}\pi^{\frac{d-1}{2}} \Gamma\left(\frac{d+1}{2}\right)} \left(\Pi_\beta\right)^{\frac{d-1}{2}} .
\end{align}
Furthermore, the integral of the terms containing the derivative of the square bracket in \eqref{eq:omegaphi42}
with respect to $|\vec{k}|$, is logarithmically divergent for large values of $|\vec{k}|$. 
In this case, it is necessary to set an upper cutoff $\mu$ on the $k$ integral. 
The $\mu$ scale is arbitrary, but one would naturally expect it to be of order $T$ at high temperature. 
Proceeding in this way, we then obtain for the free energy the result
\begin{align}
\Omega_{ring}^{d-odd}(T) &=  VT \left[ \frac{2}{d-1} + \ln \left(1+ \frac{\mu^2}{\Pi_\beta} \right)\right] \nonumber \\ &\times
\frac{(-1)^{(d+1)/2} }{2^d \pi^{(d-1)/2} \Gamma \left( \frac{d+1}{2}\right)} \left(\Pi_\beta\right)^{(d-1)/2} 
\label{eq:phi4odd}
\end{align}

Again, the factor $ \left(\Pi_\beta \right)^{(d-1)/2} $ arises on dimensional grounds.  However, since in this case $(d-1)/2$ is an integer, this factor no longer introduces a nonanalyticity in the coupling constant. Instead, in this case the nonanalyticity will only be logarithmic, due to the presence of the coupling $g^2$ in the argument of the logarithm. 

Like in the scalar theory, higher order contributions to the thermal free energy in QED are also nonanalytic in the coupling constant. 
Such contributions arise from the set of ring diagrams shown in Fig.~\ref{qedvl_rings}. In QED the photon self-energy is gauge invariant and satisfies the transversality condition:
\begin{figure}
\centering
\includegraphics[scale=0.27]{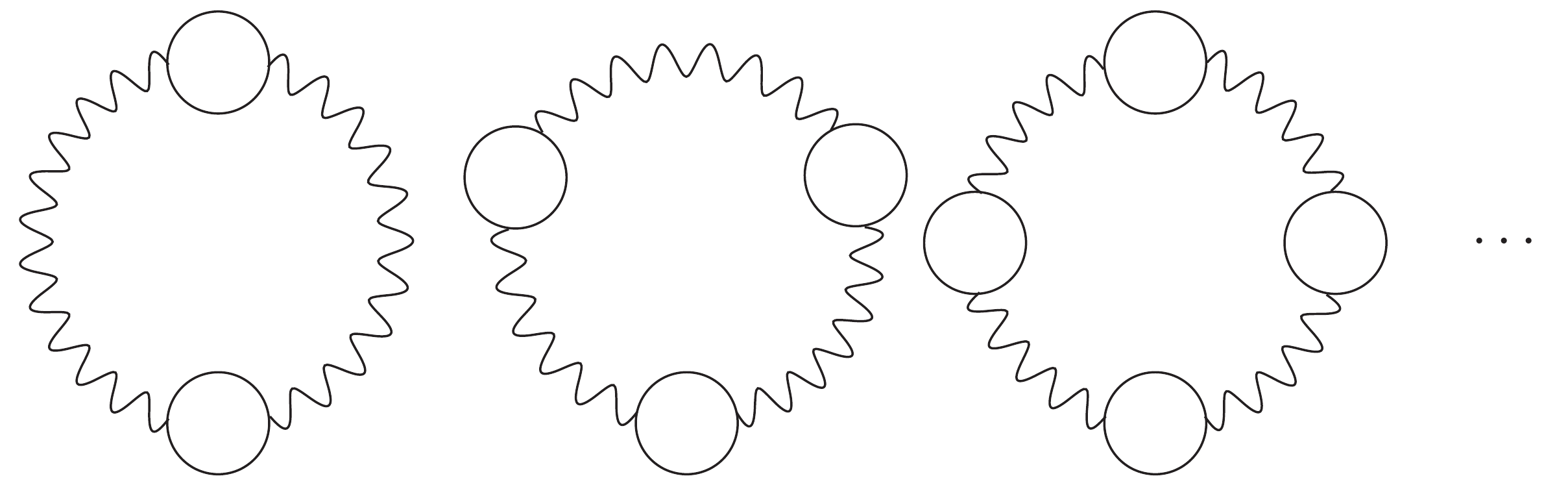}
\caption{Ring diagrams in QED. Small blobs represent one-loop photon self-energy diagrams.}
\label{qedvl_rings}
\end{figure}
\begin{align}
k^\mu \Pi_{\mu \nu}(k) =0.
\label{eq:trnas}
\end{align} 
When these self-energies are inserted in the ring loop, the result will be manifestly  gauge invariant, since the gauge-dependent part of the photon propagator vanishes 
when it multiplies $\Pi_{\mu\nu}$ as a consequence of the current conservation. Thus, we may use, without loss of generality, the Feynman gauge, where the photon propagator is rather 
similar to the propagator of the massless scalar field in $\phi^4$ theory. Due to \eqref{eq:trnas}, the photon self-energy may be expressed
in terms of two projection operators   $P^L_{\mu \nu}$ and $P^T_{\mu \nu}$, which are $d$-dimensionally transverse, the former being $(d-1)$-dimensionally longitudinal, while the latter is also $(d-1)$-dimensionally transverse.
Their components are 
\begin{subequations}
\begin{align}
P_{00}^T&=P_{0i}^T=P_{i0}^T= 0 \\
P_{ij}^T&= \delta_{ij} -\frac{k_i k_j}{\underline{k}^2} \\
P_{\mu\nu}^L&= P_{\mu\nu}^T - \frac{k_\mu k_\nu}{k^2} + \eta_{\mu\nu}  .
\end{align}
\end{subequations}
Thus, one can write the photon self-energy in the form
\begin{align}
\Pi_{\mu \nu}(k) = F(k) P^L_{\mu \nu} + G(k) P^T_{\mu \nu},
\label{eq:pibasi}
\end{align}
where $F$ and $G$ are scalar functions of $k_0$ and $|\underline{k}|$.
The potential singularities in the ring diagram occur at high temperature in the static limits $F(0, \vec{k} \rightarrow 0)$ and  $G(0, \vec{k} \rightarrow 0)$. Upon examination of Eq.~\eqref{eq:pibasi} 
and by using the QED Ward identity, one finds that:
\begin{align}
G(0, \vec{k} \rightarrow 0) &= \frac{1}{d-2} \Pi_{ii} (0, \vec{k} \rightarrow 0)\nonumber \\ & =  \frac{e^2  T}{d-2} \sum_{p_0}  \int \frac{d^{d-1} p}{(2 \pi)^{d-1}} \mathrm{Tr} \left( \gamma_i \frac{ \partial S(p)}{\partial p_i} \right) 
\nonumber \\ &=0
\end{align}
since the free electron propagator $S(p)$ vanishes for asymptotic momenta $|p_i| \rightarrow \infty$.
This result reflects the fact that the magnetic mass is zero in QED.

On the other hand, the leading high temperature behavior of $F(0,\vec{k} \rightarrow 0 )$ is:
\begin{align}\label{eq10}
F_{\beta} &= \Pi_{00}(0, \vec{k}\rightarrow 0) \nonumber \\ & =  \frac{e^2 2^{E(d/2)}(d-2)}{(2 \pi)^{d-1}}\int \frac{d^{d-1}p}{(2 \pi)^{d-1}} \frac{1}{|p|} \frac{1}{e^{|\vec{p}|/T}+1} \nonumber \\ &=  e^2 T^{d-2} \frac{(1-2^{3-d}) \Gamma(d-1) \zeta(d-2) }{2^{d-E(d/2)-2}\pi^{(d-1)/2} \Gamma\left( \frac{d-1}{2} \right)}
\end{align}
which is similar to the result \eqref{eq:pipahi} obtained in the case of the scalar field theory. 
Note that $(F_{\beta})^{1/2}$ generates an effective electric mass for the photon.

Using the above results and the properties
\begin{align}
P^{L\mu \sigma} P^L_{\sigma \nu} = P^{L \mu}_{\nu}, \,\,\,\,\,\, P^{L \mu}_{\mu} = 1,
\end{align}
we obtain the following contribution to the ring free energy in QED
\begin{align}
\Omega_{ring}(T) &= \frac{1}{2} VT \int \frac{d^{d-1} k}{(2 \pi)^{d-1}}  \left[ \ln \left( 1 + \frac{F_\beta}{\underline{k}^2} \right) \right. \nonumber \\ &- \left. \sum_{N=1}^{E(d/2)-1} \frac{(-1)^{N+1}}{N} \left( \frac{F_\beta}{\underline{k}^2} \right)^N \right] .
\end{align}
Apart from the substitution $F_\beta \rightarrow \Pi_\beta$ , this expression is completely analogous to that obtained in \eqref{eq:omegaphi42} for the massless $\phi^4$ theory. 
Hence, using the same arguments as in the previous case, we conclude that the free energy in QED is nonanalytic in the coupling $\alpha= e^2/4\pi$. 
Thus, we find that for $d$ even [see Eq. \eqref{eq:omegaphi4even}] 
\begin{align}\label{eq13}
\Omega^{d-even}_{ring}(T) & =  VT \frac{(-1)^{d/2+1}}{2^d \pi^{(d-3)/2} \Gamma \left( \frac{d+1}{2} \right)}  \left(F_\beta \right)^{(d-1)/2} ,
\end{align}
where $F_{\beta}$ is given by \eqref{eq10}.
In particular, in four-dimensional space-time, \eqref{eq13} reduces at high temperature to the well-known QED result \cite{kapusta:book89,lebellac:book96}
\begin{equation}\label{eq14}
\Omega^{d=4}_{ring}(T) =  - VT \frac{\left(F_\beta \right)^{3/2}}{12 \pi} = -\frac{VT^4}{12 \pi}\left(\frac{e^2}{3}\right)^{3/2} .
\end{equation}
On the other hand, when the dimension of space-time  is odd we obtain [compare with Eq.~\eqref{eq:phi4odd}]
\begin{align}\label{eq15}
\Omega_{ring}^{d-odd}(T)  & =  VT \left[ \frac{2}{d-1} + \ln \left(1+ \frac{\mu^2}{F_\beta} \right)\right] \nonumber \\ &\times
\frac{(-1)^{(d+1)/2} }{2^d \pi^{(d-1)/2} \Gamma \left( \frac{d+1}{2}\right)} \left(F_\beta\right)^{(d-1)/2} .
\end{align}
We note here that in Eqs. \eqref{eq13} and \eqref{eq15}, the factor $\left(F_\beta\right)^{(d-1)/2}$, which has dimensions of $(\mbox{mass})^{d-1}$, 
basically arises on dimensional grounds, ensuring that $\Omega$ has the correct dimension of energy. 
When $d$ is even, the exponent $(d-1)/2$ is a half-integer, 
which leads to a powerlike nonanalyticity of the form $(\alpha)^{(d-1)/2}$. On the other hand, when $d$ is odd, this exponent is an integer, so that 
it does not lead to a breakdown of perturbation theory.
In this case the nonanalyticity is only logarithmic, due to the presence of the coupling $\alpha$ in the argument of the logarithm.

The above nonanalyticities of the free energy are due to the infrared divergences which are generated by the thermal interactions.
The occurrence of softer nonanaliticities when $d$ is odd,
is a consequence of the fact [see the remarks following Eq. \eqref{eq:omegaphi41}]
that in this case the infrared singularities of the free energy are weaker.

\acknowledgments

We would like to thank CNPq (Brazil) for a grant.



\end{document}